\documentclass[runningheads]{llncs}
\usepackage[T1]{fontenc}
\usepackage{graphicx}
\usepackage{booktabs} 
\usepackage{array}
\usepackage{makecell}
\usepackage{hyperref}
\newcolumntype{P}[1]{>{\raggedright\arraybackslash}p{#1}}

\begin{document}
%

\title{autoPET IV challenge:\\ Incorporating organ supervision and human guidance for lesion segmentation in PET/CT}

\titlerunning{autoPET IV challenge}
%
\author{Junwei Huang\inst{1} \and Yingqi Hao \inst{1} \and Yitong Luo \inst{1} \and Ziyu Wang \inst{1} \and Mingxuan Liu \inst{1} \and Yifei Chen \inst{1} \and Yuanhan Wang \inst{1} \and Lei Xiang \inst{2} \and Qiyuan Tian\inst{1}\thanks{Corresponding author}}
\authorrunning{Junwei H. et al.}

%
\institute{School of Biomedical Engineering, Tsinghua University, Beijing, China \and
Subtle Medical Inc., Shanghai, China\\
\email{qiyuantian@tsinghua.edu.cn}
}
\maketitle              
\begin{abstract} 
Lesion Segmentation in PET/CT scans is an essential part of modern oncological workflows. To address the challenges of time-intensive manual annotation and high inter-observer variability, the autoPET challenge series seeks to advance automated segmentation methods in complex multi-tracer and multi-center settings. Building on this foundation, autoPET IV introduces a human-in-the-loop scenario to efficiently utilize interactive human guidance in segmentation tasks. In this work, we incorporated tracer classification, organ supervision and simulated clicks guidance into the nnUNet Residual Encoder framework, forming an integrated pipeline that demonstrates robust performance in a fully automated (zero-guidance) context and efficiently leverages iterative interactions to progressively enhance segmentation accuracy. Our source code is available at \href{https://github.com/huang-jw22/autoPET-4-submission/tree/master}{https://github.com/huang-jw22/autoPET-4-submission} .

\end{abstract}
\section{Introduction}
Positron Emission Tomography combined with Computed Tomography imaging (PET/CT) supplies combined metabolic and anatomical information, acting as a powerful imaging modality for clinical diagnosis and treatment planning \cite{Farwell2014,Schwenck2023}. However, manual lesion segmentation in PET/CT scans remains a time-consuming process susceptible to inter-observer variability, which motivates the development of automated methods to improve both efficiency and reproducibility. Deep learning has emerged as a powerful paradigm for automated segmentation of medical images, and its feasibility for handling diverse multi-tracer and multi-center PET/CT data has been robustly demonstrated in preceding autoPET challenges \cite{autopet_nature}.

The autoPET IV challenge extends this pursuit by investigating the role of interactive human guidance in the segmentation process. The challenge utilizes a large-scale dataset inherited from previous iterations, comprising 1014 FDG-tracer and 597 PSMA-tracer scans \cite{gatidis2022fdgpetct,jeblick2024psmapetct}. The focus this year is a human-in-the-loop segmentation scenario, where incremental human guidance is provided in the form of foreground and background clicks to simulate a clinical workflow where an automated tool is progressively refined by an expert user.

Our method builds upon the well-established nnU-Net framework \cite{isensee2021nnu}, which has consistently demonstrated high accuracy and strong generalization ability across numerous medical segmentation tasks, including prior autoPET competitions \cite{autopet_nature,kalisch2024autopetiiichallengeincorporating,rokuss2024fdg}. We took valuable experiences from previous top-performing teams, forming an integrated pipeline for this task. Various training strategies were investigated to teach the model progressively improve with incremental guidance while maintaining robustness in zero/few-guidance settings.

\section{Methods}
\subsection{Data}
\subsubsection{Training data.}
The autoPET IV dataset was used for both training and validation, including 1,014 FDG cases and 597 PSMA cases \cite{Farwell2014,Schwenck2023}. Specifically, the FDG dataset comprises 501 patients diagnosed with histologically proven malignant melanoma, lymphoma, or lung cancer, along with 513 negative control patients. The PSMA dataset includes pre- and/or post-therapeutic PET/CT images of male individuals with prostate carcinoma, encompassing images with (537) and without PSMA-avid tumor lesions (60). 

In autoPET IV, the role of human interaction is emphasized. For each case, 10 Foreground clicks and 10 Background clicks are simulated and modeled as 3D Gaussian maps using the provided script, then concatenated with the original PET/CT data to form a 4-channel input.

\subsubsection{Validation data.}
5-fold cross-validation was performed for model development and evaluation. An 80/20 train-val split was applied on the autoPET dataset to create the training data and validation data for 5 folds. 

\subsection{Data pre-processing}
Experiment planning and data pre-processing were conducted using the default planner and preprocessor of nnUNetv2 \cite{isensee2021nnu}. CT images were normalized with the "CT Normalization" scheme that conducts percentile clipping before normalization, while ZScore Normalization was applied to PET images and two clicks channel. Default data augmentation of nnUNetv2 includes Gaussian noise, random rotation, cropping, Gaussian blur, down-sampling, and gamma correction.

\subsection{Algorithm/model}
Our segmentation pipeline is built upon the robust nnUNet Residual Encoder framework \cite{isensee2024nnu}. To tailor this powerful baseline for the specific challenges, we integrated several key components, including an upstream tracer classification module, a dual-headed segmentation architecture with organ supervision, and a post-processing model based on PET SUV thresholding.

\subsubsection{Tracer Classification.}
FDG and PSMA tracers exhibit fundamentally different biodistribution patterns, which presents a significant challenge for a unified segmentation model. Training models on separate tracer datasets showed prospects in enhancing model's performance on corresponding tracer images. To achieve efficient and accurate tracer classification, We adopted the public weights from the autoPET 2024 runner-up team \cite{kalisch2024autopetiiichallengeincorporating}. Their model incorporates two separate ResNet trained on coronal and sagittal Maximum Intensity Projections, followed by a multilayer perceptron (MLP) that receives the concatenated features from the frozen backbones of both models, and outputs a binary prediction of the tracer type. We verified that this pre-trained model achieved 100\% classification accuracy across all training cases.

\subsubsection{Incorporating Human Guidance into Training Data.}
In autoPET IV challenge, each methods are evaluated in two aspects: The ability to efficiently utilize incremental human guidance information, and the ability to perform optimally in densely-guided scenarios. The way of incorporating guidance information into the training data largely affects the model's robustness and generalization ability across different amounts of guiding clicks. We mainly explored two ways of incorporating interactive information into the training process: 

\begin{itemize}
    \item \textbf{Full-Guidance Training.} This initial strategy aims to train a model that quickly adapts to human guidance and achieve maximum segmentation accuracy when provided with dense user interactions. In this approach, we concatenate all 10 foreground/background clicks (modelled as 3D Gaussian heatmaps) together with the PET/CT images to form a 4-channel input for each case (Channnel 0: CT; Channel 1: PET; Channel 2: FG Clicks; Channel 3: BG Clicks).\\
    
    \item \textbf{Stochastic Click Sampling.}
    A model trained exclusively on dense guidance often fails to establish a strong zero-guidance baseline and struggles with sparse inputs. To address this, we implemented a stochastic click sampling strategy. For each training sample loaded, we randomly sampled an integer k (from 0 to 10) according to a predefined probability distribution. The network was then provided with guidance maps generated from only the first k foreground and k background clicks. This approach exposes the model to the full spectrum of interactive scenarios, forcing it to learn a robust and flexible response to varying levels of user input.
\end{itemize}

\subsubsection{PSMA-specific Model Development.}
\label{sec:PSMA models}
Our lesion segmentation models are primarily built upon the nnUNetv2 framework, utilizing the most recent Residual Encoder UNet architecture (ResEnc-M/L) \cite{isensee2024nnu}. As reported in \cite{kalisch2024autopetiiichallengeincorporating}, separate training on the PSMA dataset achieved significant improvement in PSMA segmentation accuracy. We tested various training strategies and selected 3 most promising models for final 5-fold cross validation.
\begin{itemize}
    \item \textbf{Model V0: Weighted loss function with Dense Guidance.} The specificity and high sensitivity of PSMA in prostate carcinoma cases contributes to the appearance of numerous small and sparsely-distributed metastatic lesions \cite{bubendorf2000metastatic}, which are challenging for standard segmentation model. To guide the model place more emphasis on these small foreground lesions, we adjusted the weight of DiceCE loss from equal weights to Dice:CE = 2:1. Besides, the smoothing term is omitted in the Dice loss calculation to make training more stable, as suggested by \cite{rokuss2024fdg}. The model adopted the ResEnc-M architecture and was trained on the full-guidance dataset (10 clicks) for 1000 epochs with a patch size of [192, 192, 192] and a batch size of 3.\\
    
    \item \textbf{Model V1: Second-Stage Fine-Tuning for Interactive Performance.} The dense-guidance training would potentially result in a specialist model that is highly dependent on user input and underperform in zero- or few-click scenarios. To enhance robustness with sparse guidance, a second stage fine-tuning of model V0 was conducted on the same training data but using stochastic click sampling. The distribution was heavily skewed towards scenarios with minimal guidance (e.g., 40\% probability for 0 clicks, 20\% for 1 click) to specifically improve the model's baseline and sparse-guidance performance. This stage was run for 250 epochs with a reduced initial learning rate of 2e-4 to ensure stable adaptation without catastrophic forgetting.\\
    
    \item \textbf{Model V2: Fine-tuning pre-trained model with Balanced Stochastic Sampling.} A strong anatomical background has been proved helpful for enhancing lesion segmentation accuracy and mitigating false positive segmentation on organs with high physiological uptake \cite{rokuss2024fdg,kalisch2024autopetiiichallengeincorporating}. To provide anatomical prior knowledge, we utilized the publicly available weights from \cite{rokuss2024fdg}, which had been trained on a diverse, multi-modal medical imaging dataset. The single-channel pre-trained weights were expanded to our 4-channel input by duplicating them for the CT and PET channels and initializing the two click-guidance channels to zero. We then fine-tuned this model on the PSMA dataset for 1000 epochs using stochastic click sampling. The click sampling distribution for this model was designed to be more balanced, with significant weight on both zero-click (10\%) and dense-click (30\%) scenarios. This strategy was chosen to leverage the strong baseline prior from the pre-trained weights while progressively training the model to respond to dense interactive guidance.
 \end{itemize}

\subsubsection{Unified Multi-Tracer Model with Organ Supervision.}
For FDG dataset, however, the improvement of separate training is not significant. This conclusion is drawn both from \cite{kalisch2024autopetiiichallengeincorporating} and our empirical experiments. Besides, given that all FDG training data were acquired from UKT while all PSMA training data were from LMU, we hypothesized that training a unified model on the combined dataset would force it to learn more robust, center-invariant features, thereby improving its generalization potential.

To better cope with multi-tracer and multi-center data, Organ Supervision was implemented to guide the model form a strong anatomical understanding which is invariant of tracer types and center differences. We utilized the "autoPET3" Trainer class proposed by \cite{rokuss2024fdg}, which introduced an auxiliary organ segmentation head focusing on a set of 10 key organs, including spleen, kidneys, liver, urinary bladder, lung, brain, heart, stomach, prostate, and glands in the head region. Pseudo organ labels were created using TotalSegmentator \cite{wasserthal2023totalsegmentator}. Equal loss weighting was applied on both segmentation head.
For final unified training, we adopted the same methods (Organ Supervision + Pre-trained model), and investigated the performance with different amount of guidance incorporated in the training data:
\begin{itemize}
    \item \textbf{Model V3: Full-Guidance Training.} The integration of anatomical prior and organ supervision improved our confidence in the model's ability to handle few-clicks situations, so we first evaluated the model's performance when trained with maximal human guidance (all 10 clicks). To ensure consistency with pre-trained weights, ResEnc-L architecture was adopted with patch size [192, 192, 192] and batch size 2. Training was conducted for 1000 epochs with initial learning rate 1e-3. 
    
    \item \textbf{Model V4: Stochastic Click Sampling.} The second experiment aimed to create a more balanced model optimized for progressive human interaction. It followed the identical architecture and hyperparameters, but was trained using our stochastic click sampling strategy. The weights distribution for generating different numbers of clicks was [0.10, 0.10, 0.10, 0.08, 0.04, 0.04, 0.04, 0.04, 0.08, 0.08, 0.30] (corresponding to 0-10 clicks), emphasizing dense-guidance scenarios while maintaining the prior ability of the pre-trained model under 0/few-click circumstances.\\
 \end{itemize}

\subsection{Data post-processing}
For data post-processing, we adopted the thresholding methods from \cite{kalisch2024autopetiiichallengeincorporating} to reduce false positive volumes, by applying a tracer-specific SUV threshold to remove segmentation masks with PET values below the threshold. The thresholds were set to 1.5 for FDG cases and 1 for PSMA cases.

\subsection{Training and test parameters}
All models were trained using the nnUNetv2 framework. 3 PSMA models were trained on NVIDIA A100 GPUs, while the 2 unified models were trained on NVIDIA A800 GPUs. Specific training parameters such as patch size, batch size, epoch counts and initial learning rate for each experiment are detailed in their respective sections above. Any parameters not explicitly mentioned were kept to the nnUNetv2 default settings.

For inference, we employ test-time augmentation (TTA) to enhance prediction robustness. To comply with the challenge's time constraints, we estimate the time required for predicting the original image, and compute the number of mirrored axes allowed without exceeding a 40 second limit for each fold. The final segmentation is produced by averaging the softmax probabilities from multiple augmented predictions.

\subsection{Github repository}
Our code and trained weights are available at \href{https://github.com/huang-jw22/autoPET-4-submission/tree/master}{https://github.com/huang-jw22/autoPET-4-submission/tree/master} .

\section{Results}
\subsection{PSMA lesion segmentation model}
5-folds training has been performed for all 3 methods described in Section \ref{sec:PSMA models}. Average cross validation results are presented in Tab \ref{PSMA validation results}. It is important to note that these metrics are not directly comparable, as each model was trained on a dataset with a different click-guidance distribution. Models trained with stochastic sampling (V1, V2) were evaluated on validation sets that also contained varied click counts, which naturally results in lower average performance compared to the dense-guidance model (V0) evaluated on a dense-guidance validation set.
\begin{table*}[h]
\caption{Averaged 5-fold cross-validation results for the PSMA-specific models.}\label{PSMA validation results}
\centering
\small 
\begin{tabular*}{\textwidth}{c @{\extracolsep{\fill}} c @{\extracolsep{\fill}} ccc}
\toprule
\textbf{Version} & \textbf{Training Data} & \textbf{DICE} & \textbf{FPV} & \textbf{FNV} \\
\midrule
\renewcommand{\arraystretch}{2.5} 
\textbf{Model V0} & Full Guidance & 0.75 & 5.86 & 10.87 \\
\textbf{ Model V1} & Stochastic Sampling & 0.66 & 11.30 & 13.12 \\
\textbf{ Model V2} & Stochastic Sampling & 0.72 & 8.11 & 12.76 \\
\bottomrule
\end{tabular*}%
\vspace{0.5cm}
\end{table*}

To have a fair comparison of the 3 models and evaluate interactive abilities, we randomly chose 100 PSMA cases and generated 11 different inputs (with 0-10 clicks) for each of them. Interactive evaluation was then conducted by giving 11 predictions for all cases using the 3 models. 

Results of the evaluation are presented in Tab \ref{tab:interactive metrics}. It must be noted that the predictions were made by the 5-folds ensemble and all cases had already been seen by the models during training, so no guarantee of final test performance can be made. However, some valuable conclusions can still be yielded:

\begin{itemize}
    \item Model V0 showed terrible performance on 0/few-clicks circumstances. This confirms that training exclusively on dense guidance makes the model too dependent on user interaction and fails in sparsely-guided context.\\
    
    \item Despite high Dice score, Model V1 showed significantly higher False Positive Volume across all click counts. This observation suggests that a sudden absence of interactive information might cause the model to develop an overly aggressive prediction strategy to compromise the lack of external guidance.\\
    
    \item Model V2 demonstrated most stable performance and a significant correlation between number of clicks and segmentation quality. An increasing Dice score and decreasing FPV/FNV values indicate that the model efficiently utilized human guidance. The baseline performance with no guidance provided is also acceptable.
    
 \end{itemize}

\begin{table*}[ht]
\centering
\small
\caption{Average evaluation results of PSMA models across 0-10 clicks interaction.}
\label{tab:interactive metrics}
\begin{tabular*}{\textwidth}{c @{\extracolsep{\fill}} ccc ccc ccc}
\toprule
& \multicolumn{3}{c}{\textbf{DICE}} & \multicolumn{3}{c}{\textbf{FPV}} & \multicolumn{3}{c}{\textbf{FNV}} \\
\cmidrule(lr){2-4} \cmidrule(lr){5-7} \cmidrule(lr){8-10}
\textbf{Clicks} & \textbf{V0} & \textbf{V1} & \textbf{V2} & \textbf{V0} & \textbf{V1} & \textbf{V2} & \textbf{V0} & \textbf{V1} & \textbf{V2} \\
\midrule
0  & 0.000 & 0.708 & 0.619 & 0.000 & 13.927 & 0.991 & 190.238 & 4.033 & 5.641 \\
1  & 0.599 & 0.808 & 0.775 & 0.637 & 13.814 & 0.378 &  27.328 & 4.057 & 4.000 \\
2  & 0.723 & 0.832 & 0.816 & 0.687 & 13.382 & 0.377 &  11.061 & 3.533 & 2.801 \\
3  & 0.781 & 0.845 & 0.837 & 0.703 & 13.347 & 0.367 &   5.503 & 3.263 & 2.290 \\
4  & 0.810 & 0.847 & 0.849 & 0.707 & 13.261 & 0.365 &   4.935 & 3.139 & 2.190 \\
5  & 0.823 & 0.851 & 0.855 & 0.668 & 13.146 & 0.365 &   3.752 & 2.876 & 2.119 \\
6  & 0.839 & 0.855 & 0.860 & 0.703 & 13.159 & 0.364 &   3.205 & 2.729 & 1.987 \\
7  & 0.848 & 0.857 & 0.865 & 0.677 & 13.138 & 0.361 &   2.724 & 2.571 & 1.956 \\
8  & 0.854 & 0.859 & 0.866 & 0.667 & 12.948 & 0.352 &   2.399 & 2.390 & 1.831 \\
9  & 0.860 & 0.861 & 0.869 & 0.640 & 12.886 & \textbf{0.330} &   2.038 & 2.361 & 1.782 \\
10 & 0.864 & 0.862 & \textbf{0.871} & 0.635 & 12.868 & 0.332 &   1.823 & 2.202 & \textbf{1.613} \\
\bottomrule
\end{tabular*}%
\vspace{0.5cm}

\end{table*}

\subsection{Unified lesion segmentation model}
Due to computational constraints, the 5-fold cross-validation for these models was partially completed at the time of this analysis, with results available from three folds for Model V3 (full guidance) and two folds for Model V4 (stochastic sampling). Similar interactive evaluation was conducted on 250 FDG cases. Results of the evaluation is presented in Tab \ref{tab:fdg_model_comparison}. 

Both methods exhibited a stable 0-click performance and progressive performance gains with incremental user guidance. The results confirmed our assumptions that anatomical prior and organ supervision guarantees the model's generalization ability across different interactive levels, even if all training data was provided with full guidance (model V3). 

A closer analysis reveals a performance trade-off between the two models: Model V3 presents better performance under dense guidance scenarios, while Model V4 outperforms with sparse clicks inputs. To better leverage their individual specialty, we decided to adopt a hybrid strategy during inference, by flexibly selecting model version based on the number of input guiding clicks. Model V4 would be utilized in early interactive phase (0-4 clicks), while model V4 would be chosen in densely guided steps (5-10 clicks). This approach ensures that the optimal model is deployed to achieve better precision in different interactive scenarios.

\begin{table*}[ht]
\centering
\small
\caption{Average evaluation results of unified models across 0-10 clicks interaction.}
\label{tab:fdg_model_comparison}
\begin{tabular*}{\textwidth}{c @{\extracolsep{\fill}} cc cc cc}
\toprule
& \multicolumn{2}{c}{\textbf{DICE}} & \multicolumn{2}{c}{\textbf{FPV}} & \multicolumn{2}{c}{\textbf{FNV}} \\
\cmidrule(lr){2-3} \cmidrule(lr){4-5} \cmidrule(lr){6-7}
\textbf{Clicks} & \textbf{V3} & \textbf{V4} & \textbf{V3} & \textbf{V4} & \textbf{V3} & \textbf{V4} \\
\midrule
0  & 0.739 & 0.788 & 0.694 & 0.651 & 8.869 & 5.958 \\
1  & 0.811 & 0.837 & 0.647 & 0.500 & 7.040 & 5.631 \\
2  & 0.844 & 0.853 & 0.660 & 0.469 & 5.728 & 4.420 \\
3  & 0.861 & 0.862 & 0.700 & 0.459 & 4.459 & 3.874 \\
4  & 0.869 & 0.865 & 0.678 & 0.424 & 3.805 & 3.310 \\
5  & 0.875 & 0.870 & 0.684 & 0.422 & 2.798 & 2.537 \\
6  & 0.880 & 0.873 & 0.688 & \textbf{0.418} & 2.630 & 2.249 \\
7  & 0.883 & 0.873 & 0.703 & 0.435 & 2.411 & 2.090 \\
8  & 0.886 & 0.875 & 0.714 & 0.446 & 2.065 & 1.934 \\
9  & 0.888 & 0.876 & 0.721 & 0.450 & 1.805 & 1.848 \\
10 & \textbf{0.889} & 0.877 & 0.725 & 0.434 & \textbf{1.589} & 1.658 \\
\bottomrule
\end{tabular*}%
\vspace{0.5cm}
\end{table*}

\section{Final Submission}
For final submission, we followed the described pipeline, integrating a tracer classifier, separate models for different tracer types, with SUV thresholding as final post-processing. For PSMA cases, we submitted Model V2 (fine-tuning pre-trained model with balanced stochastic sampling), which exhibited best interactive performance during evaluation; for FDG cases, we adopted the aforementioned hybrid strategy, flexibly utilizing two different models based on different scenarios. Training parameters and algorithm details have been described above and also summarized in Tab. \ref{tab4}.

\section{Conclusion}
In this work, we developed and validated a comprehensive pipeline for interactive lesion segmentation in multi-tracer and multi-center PET/CT data. The proposed pipeline integrates several key methodologies, including tracer classification, unified \& tracer-specific model training, organ supervision and PET SUV thresholding. We also explored various ways of incorporating human guidance into training data. Our experiments yielded two critical insights. First, we showed that a carefully designed stochastic sampling curriculum can effectively enhance the model's generalization ability across different guidance level. Second, we demonstrated the role of introducing anatomical prior and organ supervision in securing a strong non-guiding baseline precision while reaching a high-performance ceiling with dense interactions. Our pipeline exhibits robust performance with sparse human guidance and a proficient ability to leverage incremental human feedback to refine segmentation accuracy.

\begin{table}[ht]
\caption{Algorithm details}\label{tab4}

\begin{tabular}{P{0.2\textwidth}P{0.2\textwidth}P{0.2\textwidth}P{0.2\textwidth}P{0.2\textwidth}} 
\toprule
\textbf{Team name} & \textbf{algorithm name} & \textbf{data pre-processing} & \textbf{data post-processing} & \textbf{training data augmentation} \\
\midrule
BIRTH & BIRTH final submission &  nnUNet Preprocessor & PET SUV Thresholding & nnUNet DA scheme (Gaussian noise, random rotation, cropping, Gaussian blur, down-sampling, and gamma correction)\\
\bottomrule
\end{tabular}

\vspace{2em}

\begin{tabular}{P{0.2\textwidth}P{0.2\textwidth}P{0.2\textwidth}P{0.2\textwidth}P{0.2\textwidth}} 
\toprule
\textbf{test time augmentation} & \textbf{ensembling} & \textbf{standardized framework}  & \textbf{network architecture} & \textbf{loss} \\
\midrule
Mirroring & multi-folds cross-validation & nnUNetv2 (3D) & ResEnc UNet (3D) & DSC + CE \\
& & & & \\
\bottomrule
\end{tabular}

\vspace{2em}

\begin{tabular}{P{0.2\textwidth}P{0.2\textwidth}P{0.2\textwidth}P{0.2\textwidth}P{0.2\textwidth}} 
\toprule
\textbf{training data} & \textbf{data/model dimensionality and size} & \textbf{use of pre-trained models} & \textbf{GPU hardware for training}\\
\midrule
1014 FDG + 597 PSMA PET-CT of autoPET & 3D: 192x192x192 & Public available pre-trained models & 5x Nvidia A100 1x Nvidia A800\\
& & & & \\
\bottomrule
\end{tabular}
\end{table}


\begin{credits}
\subsubsection{\ackname} This work was supported by Tsinghua University Initiative Scientific Research Program (Student Academic Research Advancement Program).

\subsubsection{\discintname} The authors have no competing interests to declare that are relevant to the content of this article. 
\end{credits}
%
%
%
%
\bibliographystyle{splncs04}
\bibliography{ref}

\end{document}